\documentclass[twocolumn,aps,prb,amsmath,floatfix]{revtex4}
\usepackage{graphicx}
\begin{document}
\title{Correlated Dirac Fermions on the Honeycomb Lattice \\ 
studied within Cluster Dynamical Mean Field Theory}
\author{Ansgar Liebsch}
\affiliation{Institut f\"ur Festk\"orperforschung, 
             Forschungszentrum J\"ulich, 
             52425 J\"ulich, Germany}
 \begin{abstract} 
The role of non-local Coulomb correlations in the honeycomb lattice is 
investigated within cluster dynamical mean field theory combined with 
finite-temperature exact diagonalization. 
The paramagnetic semi-metal to insulator transition is found to be in 
excellent agreement with finite-size determinantal Quantum Monte Carlo 
simulations and with cluster dynamical mean field calculations based on 
the continuous-time Quantum Monte Carlo approach. As expected, the critical 
Coulomb energy is much lower than within a local or single-site formulation. 
Short-range correlations are shown to give rise to a pseudogap and 
concomitant non-Fermi-liquid behavior within a narrow range below the 
Mott transition.
\\
%
\end{abstract}
\maketitle

\section{Introduction}

The recent discovery of graphene\cite{novoselov} has greatly stimulated 
the study of the electronic properties of the honeycomb lattice.\cite{castro} 
In view of the vanishing density of states at the Fermi level, an issue 
of particular interest is the influence of electron-electron interactions.
Gonz\'alez {\it et al.}\cite{gonzalez} performed renormalization group 
calculations and showed that the suppression of screening of the long-range 
Coulomb interaction gives rise to deviations from conventional Fermi-liquid 
behavior. Lattice field theory simulations\cite{drut} indicated a Coulomb 
driven second-order semi-metal to insulator transition. 
Meng {\it et al.}\cite{meng} performed extensive variational 
Quantum Monte Carlo (QMC) simulations for the Hubbard model with varying 
cluster sizes and identified a spin-liquid phase between the semi-metallic 
state characterized by massless Dirac fermions and an antiferromagnetically 
ordered Mott insulator. The onset of the
long-range antiferromagnetic order was found to be consistent with 
previous QMC calculations for finite-size clusters.\cite{sorella,paiva}
The Mott transition of the honeycomb lattice was also investigated
\cite{jafari,tran} within single-site dynamical mean field theory \cite{dmft} 
(DMFT). However, because of the small number of nearest neighbors, 
the neglect of spatial correlations in this system is questionable and 
gives rise to a significant overestimate of the range of the semi-metallic 
behavior up to large values of the onsite Coulomb interaction. 
To account for non-local correlations in the honeycomb lattice, 
Wu {\it et al.}\cite{wu} recently applied a cluster extension\cite{kotliar} 
of DMFT (CDMFT) by using continuous-time QMC \cite{rubtsov} as impurity solver.
The transition between the semi-metallic and insulating phases was found
to occur at a considerably smaller critical Coulomb energy than within the
single-site DMFT and to be in good agreement with the variational QMC 
results by Meng {\it et al.}\cite{meng}        
    
In the present work we use finite-temperature exact diagonalization\cite{ed} 
(ED) in combination with cluster DMFT to investigate the two-dimensional Hubbard 
model on the honeycomb lattice for unit cells consisting of six sites.
The focus is on the dynamical properties of the non-local self-energy 
which have not been studied before. Moreover, 
in view of the large size of this unit cell and the approximate nature of 
quantum impurity solvers, CDMFT results obtained within complementary schemes 
are clearly desirable. An important advantage of ED is the accessibility of 
large Coulomb energies and low temperatures, and the absence of sign problems. 
Also, in contrast to finite-size variational QMC, ED is applicable away from 
half-filling.
On the other hand, due to the exponential growth of the Hilbert space, the 
number of levels representing the bath surrounding the cluster is severely 
limited. Here, we use 12 levels in total, i.e., six impurity levels and 
six bath levels. Since these bath states are coupled indirectly via the
onsite Coulomb repulsion within the six atom cluster, the spacing between 
excitation energies is very small. Finite-size errors are thereby greatly 
reduced, even at low temperatures, so that self-energies and spectral 
functions can be evaluated reliably at rather low real frequencies. 

The ED/CDMFT results discussed below reveal a continuous Mott transition 
in excellent correspondence with the variational QMC simulations by Meng 
{\it et al.}\cite{meng} and with the QMC/CDMFT calculations by Wu {\it et al.} 
\cite{wu} The critical onsite Coulomb energy is considerably smaller than 
the one found in single-site DMFT calculations.\cite{jafari,tran} 
Furthermore, short-range correlations included within CDMFT are shown to 
give rise to metallic and insulating contributions to the self-energy at 
the Dirac points in the Brillouin Zone, where the former dominate at low 
Coulomb interactions, and the latter increase essentially quadratically 
with the nearest-neighbor non-local self-energy component. These terms 
lead to the excitation gap above the Mott transition. Below the transition  
they yield a narrow pseudogap.
Thus, short-range correlations induced
via onsite Coulomb repulsion give rise to deviations from Fermi-liquid 
behavior in some range below the critical interaction strength. Also,
the effective mass enhancement does not diverge at the Mott transition,
but increases to a finite value. The opening of the pseudogap below the 
transition, and the variation of the effective mass with Coulomb energy,
are qualitatively similar to analogous results obtained within cluster
DMFT calculations for the square lattice.\cite{kyung,zhang,park}   
 
The outline of this paper is as follows: In the next section we briefly
outline the application of ED/CDMFT to the Hubbard model for the honeycomb 
lattice. Section III provides the discussion of the results. The summary 
is presented in section IV. The focus in this work is on the paramagnetic 
semi-metal to insulator transition. Spin liquid and antiferromagnetic phases
will be addressed in a subsequent publication.   

\section{Cluster ED/DMFT for Honeycomb lattice}

In this section we discuss the combination of cluster DMFT with  
finite-temperature ED for the purpose of evaluating the effect of non-local 
Coulomb interactions on the honeycomb lattice. The Hubbard Hamiltonian 
is given by 
\begin{equation}
H=-t \sum_{\langle ij\rangle\sigma} ( c^+_{i\sigma} c_{j\sigma} + {\rm H.c.}) 
                     + U \sum_i n_{i\uparrow} n_{i\downarrow} ,
\end{equation}
where the sum in the first term includes only nearest neighbors, 
$t$ is the hopping matrix element, and $U$ is the onsite Coulomb repulsion.
The band dispersion for the honeycomb lattice may be written as 
$\epsilon({\bf k})= \pm t\vert 1+e^{ik_x\sqrt{3}}+e^{i(k_x\sqrt{3}+k_y3)/2}\vert$.
In the following we define $t=1$ as energy unit.

Let us divide the two-dimensional lattice into clusters consisting of six 
sites. Within the unit cell, the positions are specified as 
${\bf a}_1=(0,0)$, ${\bf a}_2=(1,0)$, ${\bf a}_3=(\sqrt3/2,3/2)$, 
${\bf a}_4=(\sqrt3,1)$, ${\bf a}_5=(\sqrt3,0)$, and ${\bf a}_6=(\sqrt3/2,-1/2)$. 
The nearest neighbor spacing is taken to be $a=1$. 
The supercell lattice vectors are then given by $A_{1/2}=(3\sqrt3/2,\pm3/2)$.
Within CDMFT \cite{kotliar} the interacting lattice Green's function in 
the cluster site basis is given  by
\begin{equation}
     G_{ij}(i\omega_n) = \sum_{\bf k} \left[ i\omega_n + \mu - t({\bf k})- 
                   \Sigma(i\omega_n)\right]^{-1}_{ij} ,
\label{G}
\end{equation}
where $\omega_n=(2n+1)\pi T$ are Matsubara frequencies and $\mu$ is the 
chemical potential. The ${\bf k}$ sum extends over the reduced Brillouin Zone,
$t({\bf k})$ denotes the hopping matrix for the superlattice, and  
$\Sigma_{ij}(i\omega_n)$ represents the cluster self-energy matrix in the 
site representation. The diagonal elements of the symmetric matrix $G_{ij}$ 
are identical and there are three independent off-diagonal elements: 
$G_{12}=G_{16}$, $G_{13}=G_{15}$ and $G_{14}$. Because of these symmetry 
properties, it is convenient to go over to a diagonal ``molecular orbital
basis'', in which the elements $G_m(i\omega_n)$ are given by
\begin{eqnarray}
 G_{1/2} &=& (G_{11}+2G_{13}) \pm (G_{14}+2G_{12}) \nonumber \\ 
 G_{3/4} &=& (G_{11}- G_{13}) \pm (G_{14}-G_{12})            \\ 
         &=& G_{5/6} . \nonumber
\label{Gk}
\end{eqnarray}

\begin{figure}  [t!] 
\begin{center}
\includegraphics[width=4.5cm,height=6.5cm,angle=-90]{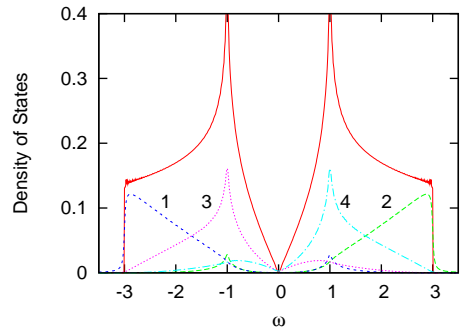}
\end{center}
\caption{(Color online)
Total density of states $\rho(\omega)$ (solid curve) of honeycomb lattice 
and cluster components $\rho_m(\omega)$ (dashed curves) within diagonal 
molecular orbital basis. For clarity, these components are divided by 
$n_c=6$.  Orbitals 3 and 4 are doubly degenerate.
$\omega=0$ defines the Fermi energy for half-filling. 
}\label{dos}\end{figure}

The self-energy matrix satisfies the same symmetry properties as $G$ and 
can therefore be diagonalized in the same fashion. These elements will be 
denoted as $\Sigma_m(i\omega_n)$.
Below we focus on the special case of half-filling. Since the density of 
states is then particle-hole symmetric with respect to $\omega=0$, 
$G_{ii}(i\omega_n)$ is purely imaginary. The same applies to $G_{13}$,
whereas $G_{12}$ and $G_{14}$ are real, corresponding to odd density of 
states components. Thus, the diagonal molecular orbital components of $G$ 
satisfy $G_2=-G_1^*$ and $G_4=-G_3^*$. Figure~\ref{dos} illustrates the 
uncorrelated density of states components in the diagonal basis, where 
$\rho_m(\omega) = -\frac{1}{\pi}\,{\rm Im}\, G_m(\omega)$ for $\Sigma=0$.
The even and odd onsite and intersite components may be obtained by inverting
Eq.~(\ref{Gk}).

A central feature of DMFT is that, to avoid double-counting of Coulomb 
interactions in the quantum impurity calculation, the self-energy must be
removed from the small cluster in which correlations are treated explicitly. 
This removal yields the impurity Green's function 
\begin{equation}
         G_0(i\omega_n) = [G(i\omega_n)^{-1} + \Sigma(i\omega_n)]^{-1} .  
           \label{G0}
\end{equation} 
For the purpose of perfoming the ED calculation we now project the diagonal 
components of $G_0(i\omega_n)$ onto those of a larger cluster consisting of 
six impurity levels and six bath levels, i.e., $n_s=12$ is the 
total number of levels. Thus,
\begin{eqnarray}
 G_{0,m}(i\omega_n) &\approx&  G^{cl}_{0,m}(i\omega_n) \nonumber\\
    &=&    \left( i\omega_n + \mu -\epsilon_m -
 \sum_{k=7}^{12} \frac{\vert V_{mk}\vert^2}{i\omega_n - \epsilon_k}\right)^{-1},
   \label{G0m}
\end{eqnarray}
where $\epsilon_m$ denotes impurity levels, $\epsilon_k$ bath levels, 
and $V_{mk}$ hybridization matrix elements. The incorporation of the 
impurity level $\epsilon_m$ in the fitting procedure yields a more accurate  
representation of $G_{0,m}(i\omega_n)$ than by projecting only onto bath 
levels. 

Assuming independent baths for the diagonal cluster molecular orbitals, each 
component $G_{0,m}(i\omega_n)$ is fitted using three parameters: one impurity 
level $\epsilon_m$, a bath level $\epsilon_k$ and a hopping integral $V_{mk}$. 
To evaluate the finite-temperature interacting Green's function of the 
cluster it is useful to transform the impurity orbitals back to the site
representation in which the Coulomb interaction is diagonal. We denote 
this transformation by $T$, where the matrix elements are given by 
\begin{eqnarray}
     T_{im} &=& \left( \begin{array}{rrrrrr}
\frac{1}{\sqrt6}& \frac{1}{\sqrt6}& 0           & \frac{1}{ \sqrt3}& \frac{1}{ \sqrt3} & 0           \\
\frac{1}{\sqrt6}&-\frac{1}{\sqrt6}&-\frac{1}{2} & \frac{1}{2\sqrt3}&-\frac{1}{2\sqrt3} & \frac{1}{2}\\
\frac{1}{\sqrt6}& \frac{1}{\sqrt6}& \frac{1}{2} &-\frac{1}{2\sqrt3}&-\frac{1}{2\sqrt3} & \frac{1}{2}\\
\frac{1}{\sqrt6}&-\frac{1}{\sqrt6}& 0           &-\frac{1}{ \sqrt3}& \frac{1}{ \sqrt3} & 0           \\
\frac{1}{\sqrt6}& \frac{1}{\sqrt6}&-\frac{1}{2} &-\frac{1}{2\sqrt3}&-\frac{1}{2\sqrt3} &-\frac{1}{2}\\
\frac{1}{\sqrt6}&-\frac{1}{\sqrt6}& \frac{1}{2} & \frac{1}{2\sqrt3}&-\frac{1}{2\sqrt3} &-\frac{1}{2}\\
                                  \end{array} \right)  . \label{T}
\end{eqnarray}
Thus, the diagonal $6\times6$ subblock of the cluster Hamiltonian, 
$h_b=(\epsilon_k\delta_{kk'})$, representing the bath levels remains unchanged, 
while the $6\times6$ impurity subblock becomes nondiagonal in the cluster 
site basis:
\begin{eqnarray}
      h_c &=& \left( \begin{array}{llllll}
             \epsilon & \tau     & \tau'    & \tau''    & \tau'    & \tau     \\
             \tau     & \epsilon & \tau     & \tau'     & \tau''   & \tau'    \\ 
             \tau'    & \tau     & \epsilon & \tau      & \tau'    & \tau''   \\  
             \tau''   & \tau'    & \tau     & \epsilon  & \tau     & \tau'     \\  
             \tau'    & \tau''   & \tau'    & \tau      & \epsilon & \tau     \\  
             \tau     & \tau'    & \tau''   & \tau'     & \tau     & \epsilon \\  
                                             \end{array} \right) , 
\end{eqnarray}
with
\begin{eqnarray}
  \epsilon &=& [(\epsilon_1+\epsilon_2) + 2(\epsilon_3+\epsilon_4)]/6 \nonumber \\
  \tau     &=& [(\epsilon_1-\epsilon_2) -  (\epsilon_3-\epsilon_4)]/6 \nonumber \\     
  \tau'    &=& [(\epsilon_1+\epsilon_2) -  (\epsilon_3+\epsilon_4)]/6 \nonumber \\  
  \tau''   &=& [(\epsilon_1-\epsilon_2) + 2(\epsilon_3-\epsilon_4)]/6  .            
\end{eqnarray}
We point out that the hopping element $t$ of the original lattice Hamiltonian 
does not appear since it is absorbed into $\tau$ via the molecular orbital 
cluster levels $\epsilon_m$ which are adjusted to fit $G_{0,m}(i\omega_n)$. 
The procedure above therefore includes not only hopping between cluster and bath. 
It also introduces four new parameters within the six-site cluster: the onsite
level $\epsilon$, and up to third-neighbor hopping parameters:  
$\tau$, $\tau'$, and $\tau''$.      
At half-filling, $\epsilon_2=-\epsilon_1$ and  $\epsilon_4=-\epsilon_3$ for
symmetry reasons, so that $\epsilon=\tau'=0$. 
In this mixed site-molecular orbital basis, the hybridization matrix elements 
$V_{mk}$ between cluster and bath molecular orbitals introduced in Eq.~(\ref{G0m}) 
are transformed to new hybridization matrix elements between cluster sites 
$i$ and bath orbitals $k$. They are given by
\begin{equation}
       V'_{ik} = (T V)_{ik} = \sum_m T_{im} V_{mk}\ .
\end{equation}
The single-particle part of the cluster Hamiltonian now reads
\begin{eqnarray}
      h_0 &=& \left( \begin{array}{ll}
                    h_c      & V'    \\
                    V'^t     & h_b   \\   \end{array} \right) . 
\end{eqnarray}
Adding the onsite Coulomb interactions to this Hamiltonian, the non-diagonal 
interacting cluster Green's function at finite $T$ can be derived from the 
expression\cite{perroni,luca}
\begin{eqnarray}
 G^{cl}_{ij}(i\omega_n) &=& \frac{1}{Z} \sum_{\nu\mu}\,e^{-\beta E_\nu}\, 
          \Big(\frac{\langle\nu\vert c_{i\sigma}  \vert\mu\rangle 
                     \langle\mu\vert c_{j\sigma}^+\vert\nu\rangle}
                                  {E_\nu - E_\mu + i\omega_n}            \nonumber\\
       &&\hskip9mm + \ \ \frac{\langle\nu\vert c_{i\sigma}^+\vert\mu\rangle 
                               \langle\mu\vert c_{j\sigma}  \vert\nu\rangle}
                                  {E_\mu - E_\nu + i\omega_n} \Big) ,  
     \label{Gcl}
\end{eqnarray}
where $E_\nu$ and $|\nu \rangle$  denote the eigenvalues and eigenvectors of 
the Hamiltonian, $\beta=1/T$ and $Z=\sum_\nu {\rm exp}(-\beta E_\nu)$ is the 
partition function. Further details concerning the evaluation of the cluster
Green's function can be found in Ref.~\onlinecite{tong} where the analogous 
procedure is discussed for the square lattice.
Since $G^{cl}_{ij}$ satisfies the same symmetry properties as $G_{ij}$, 
it is diagonal within the molecular orbital basis, with elements $G^{cl}_{m}$.
The diagonal cluster self-energy components are then given by an 
expression analogous to Eq.~(\ref{G0}):
\begin{equation}
\Sigma^{cl}_{m}(i\omega_n) = 1/G^{cl}_{0,m}(i\omega_n)-1/G^{cl}_{m}(i\omega_n) .
\label{Scl}
\end{equation} 

The important assumption in DMFT is now that this impurity cluster self-energy 
is a physically reasonable representation of the lattice self-energy. Thus, 
\begin{equation}
     \Sigma_{m}(i\omega_n) \approx \Sigma^{cl}_{m}(i\omega_n) ,  
\label{S}
\end{equation}
where, at real frequencies, $\Sigma_{m}(\omega)$ is continuous whereas
$\Sigma^{cl}_{m}(\omega)$ is discrete.

In the next iteration step, these diagonal self-energy components are 
used as input in the lattice Green's function Eq.~(\ref{G}), which in the 
molecular orbital basis may be written as
\begin{equation}
   G_{m}(i\omega_n) = \sum_{\bf k} \left[ i\omega_n + \mu - T t({\bf k}) T^{-1} 
                 -  \Sigma(i\omega_n)\right]^{-1}_{mm} ,
\label{Gm}
\end{equation}
where $T$ is the transformation defined in  Eq.~(\ref{T}). Note that 
$T t({\bf k}) T^{-1}$ is not diagonal at general ${\bf k}$ points. 
As a result, all molecular orbital components of $\Sigma(i\omega_n)$ 
contribute to all components $G_{m}(i\omega_n)$. We also point out that,
to get adequate resolution at low frequencies, because of the vanishing 
density of states, a sufficiently large number of ${\bf k}$ values near 
the Dirac points must be included in the Brillouin Zone integration.

\begin{figure}  [t!!] 
\begin{center}
\includegraphics[width=4.5cm,height=6.5cm,angle=-90]{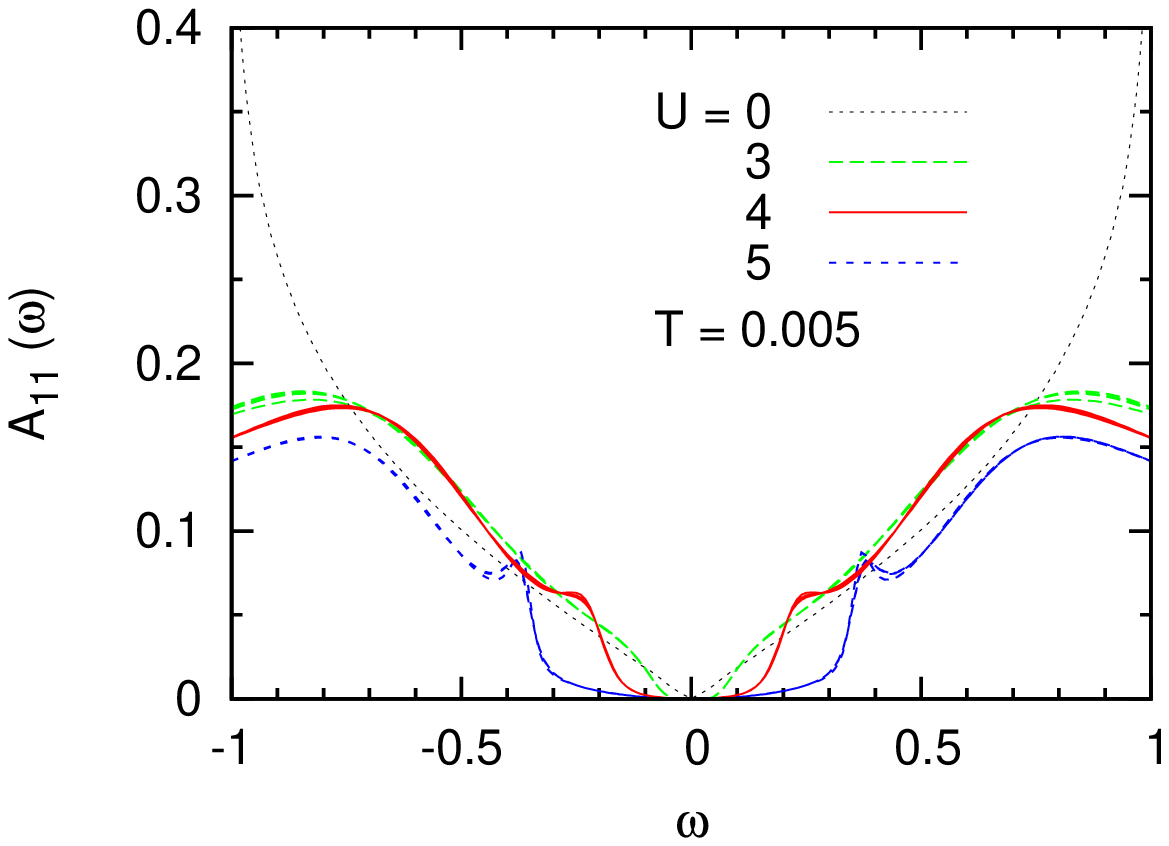}
\end{center}
\vskip-7mm \ \ \ (a)\hfill \mbox{\hskip5mm}\vskip-3mm
\begin{center}
\includegraphics[width=4.5cm,height=6.5cm,angle=-90]{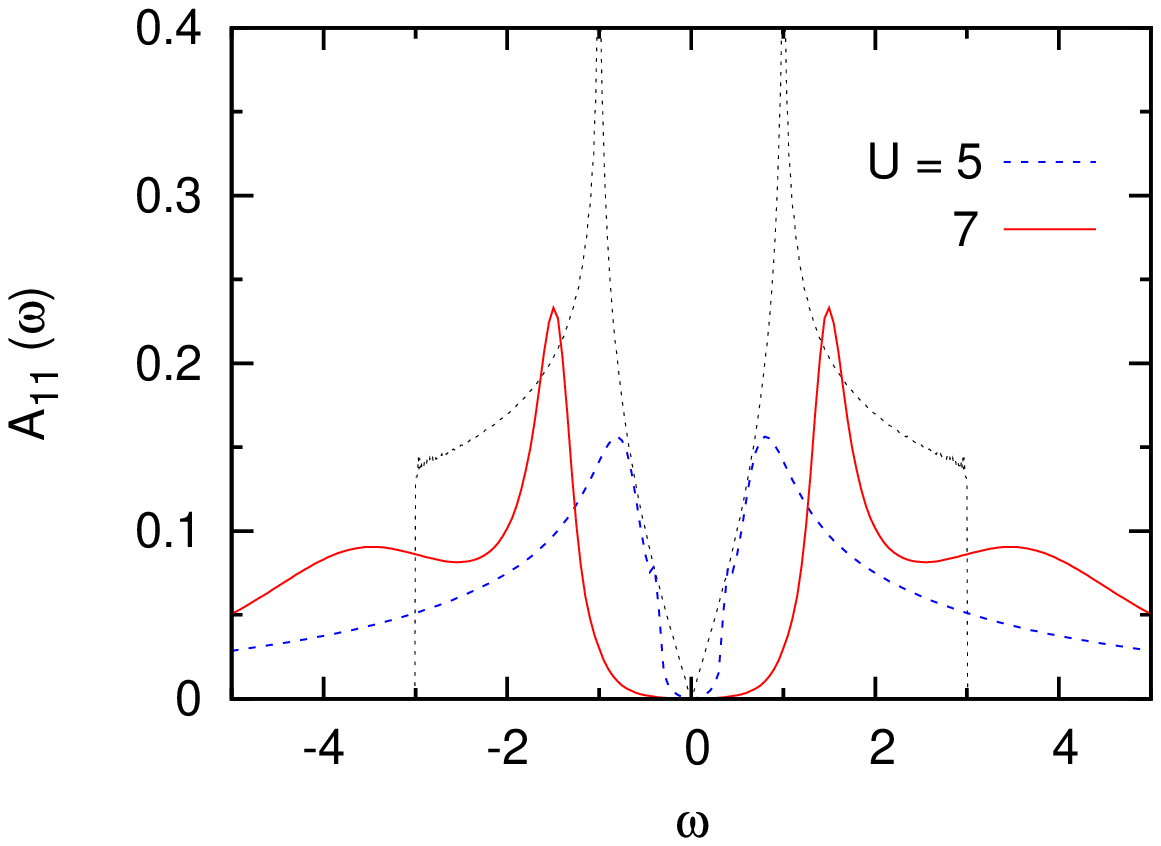}
\end{center}
\vskip-7mm \ \ \ (b)\hfill \mbox{\hskip5mm}
\caption{(Color online)
(a) Low-energy region of density of states 
$A_{11}(\omega)= -\frac{1}{\pi}\,{\rm Im}\,G_{11}(\omega)$
of honeycomb lattice for several Coulomb energies at $T=0.005$. 
The noninteracting density of states is indicated by the black dotted curve.  
Between 50 and 200 Matsubara points are used to extrapolate the lattice
Green's function to real frequencies.  
(b) Density of states over wider energy range for $U=5$ and $U=7$.
}\label{A11}\end{figure}

\section{Results and Discussion}

Figure 2(a) shows the low-energy region of the interacting density of states  
for several Coulomb energies, at temperature $T=0.005$.
These distributions are derived from an  
extrapolation of the local lattice Green's function $G_{11}(i\omega_n)$
to real frequencies. To illustrate the stability of this extrapolation, 
at each value of $U$ several curves are plotted for 50 to 200 Matsubara 
points, with an additional small energy broadening of the order of 
$0.1\omega^2$. (For $\vert\omega\vert>1$ the broadening is kept constant
at $0.1$.) At $U=3$, a tiny gap or pseudogap is seen which is near the 
limit of what can be resolved within ED/DMFT. At $U=4$, a full gap of 
width $\Delta\approx 0.25$ has opened. Its width increases approximately 
to $\Delta\approx 0.6$ when the Coulomb energy is increased to $U=5$.
This trend is consistent with the one found in Refs.~\onlinecite{meng,wu}.
The variation of the gap over a wider range of $U$ is indicated in Fig.~3(a). 
The spectral distributions in Fig.~2 show that the van Hove singularity 
at $\omega=\pm1$ is strongly broadened and its weight is shifted to lower 
energies. Above the transition, the Hubbard bands are difficult to resolve 
as long as $U$ is less than the band width, but they become pronounced 
once $U>W$, as shown in Fig.~2(b) for $U=7$.

\begin{figure}  [t!] 
\begin{center}
\includegraphics[width=4.0cm,height=6.5cm,angle=-90]{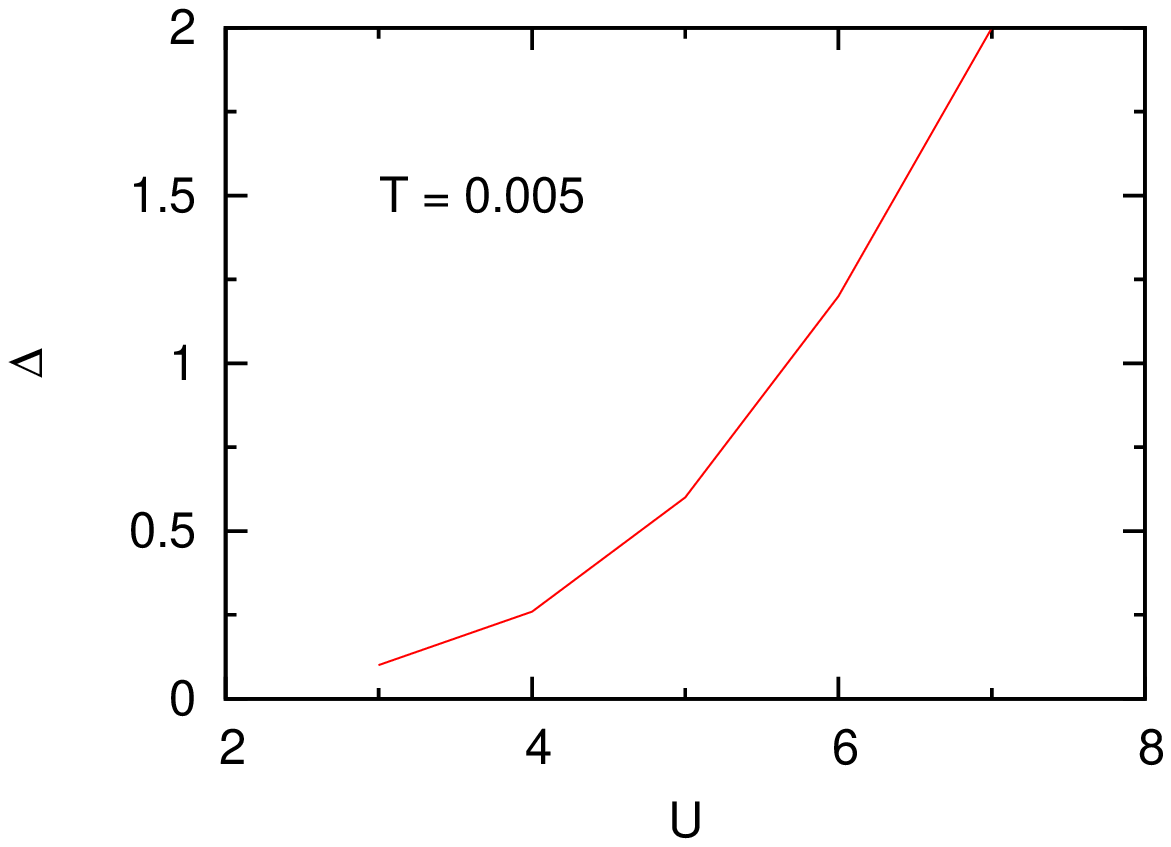}
\end{center}
\vskip-7mm \ \ \ (a)\hfill \mbox{\hskip5mm}\vskip-3mm
\begin{center}
\includegraphics[width=4.0cm,height=6.5cm,angle=-90]{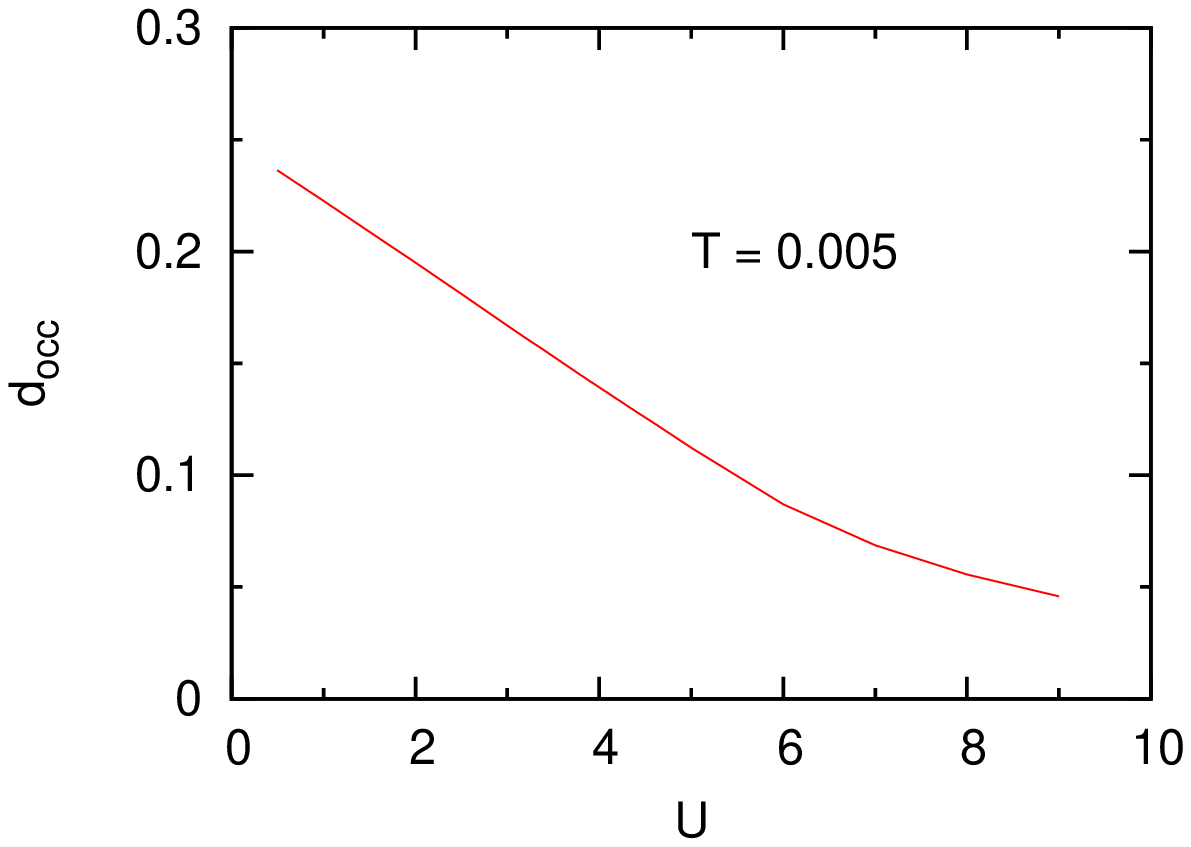}
\end{center}
\vskip-7mm \ \ \ (b)\hfill \mbox{\hskip5mm}\vskip-3mm
\begin{center}
\includegraphics[width=4.0cm,height=6.5cm,angle=-90]{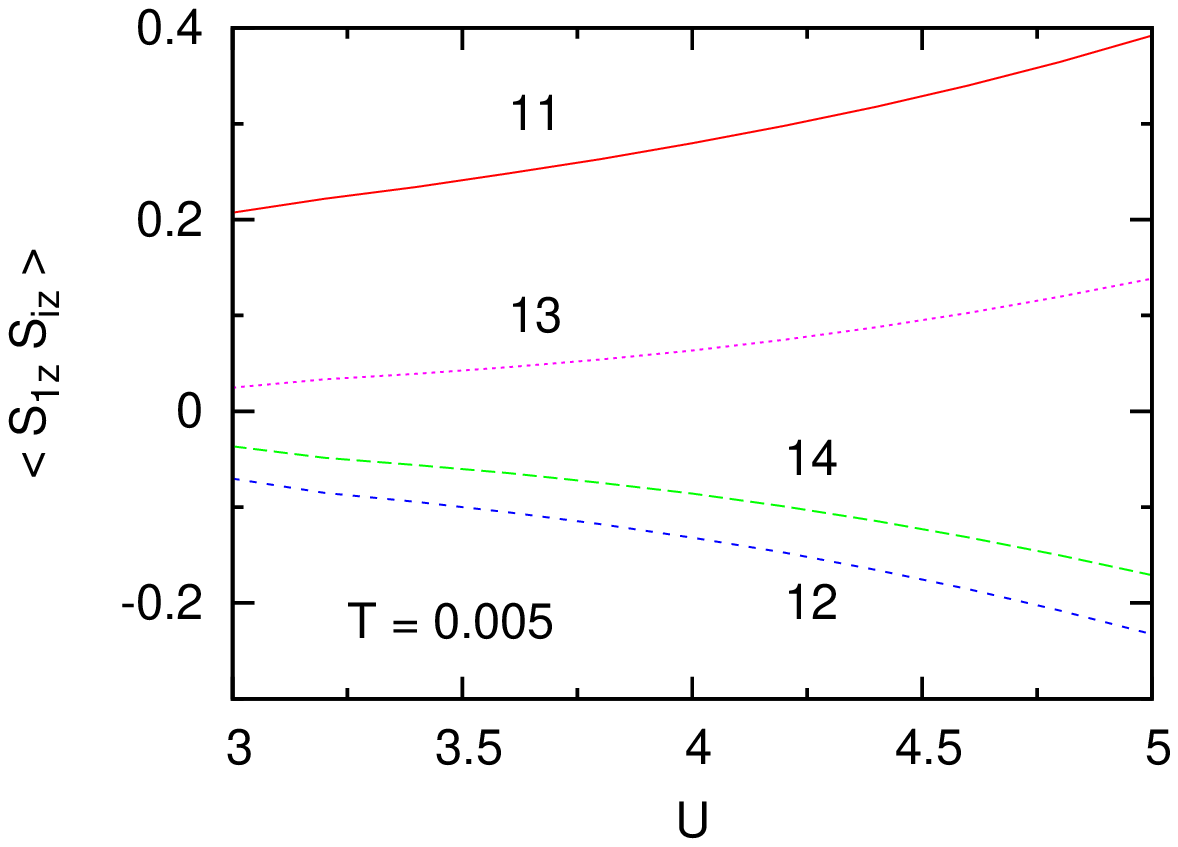}
\end{center}
\vskip-7mm \ \ \ (c)\hfill \mbox{\hskip5mm}\vskip-3mm
\begin{center}
\includegraphics[width=4.0cm,height=6.5cm,angle=-90]{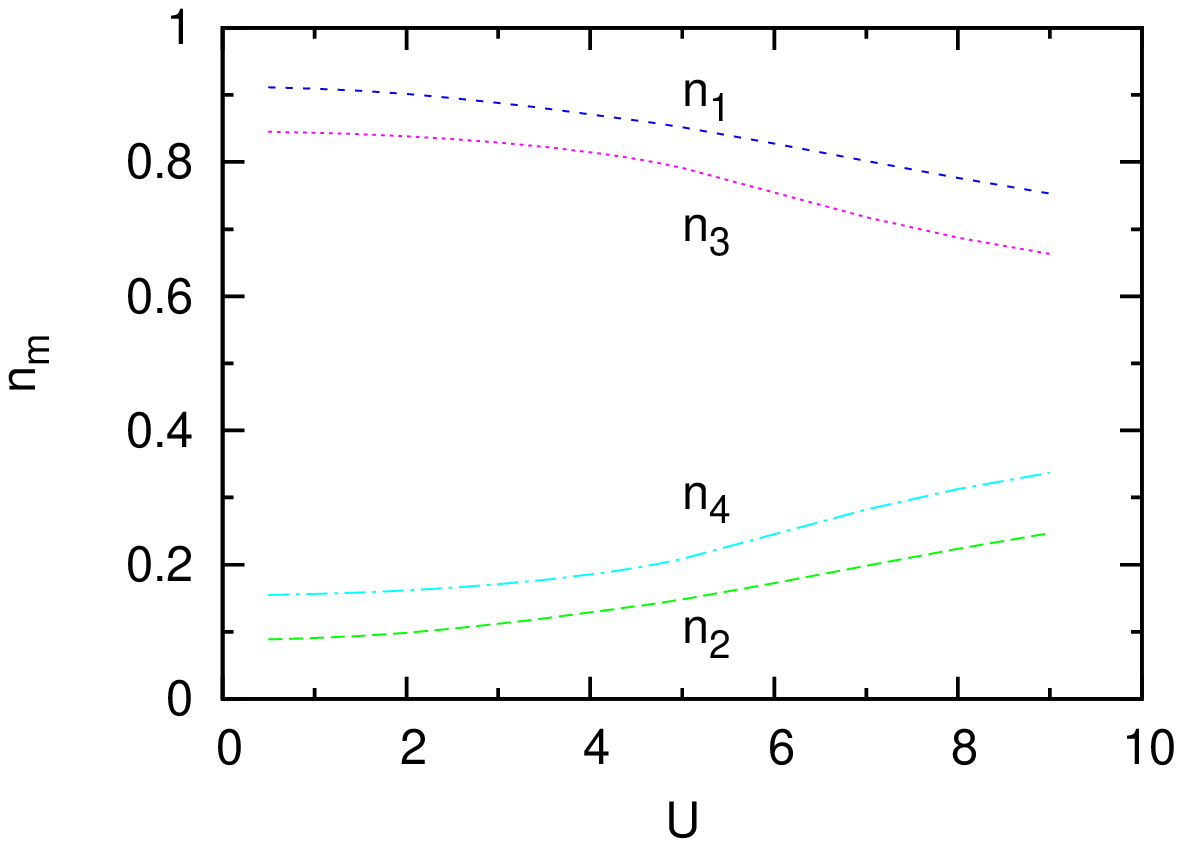}
\end{center}
\vskip-7mm \ \ \ (d)\hfill \mbox{\hskip5mm}
\caption{(Color online)
(a) Excitation gap $\Delta$, (b) average double occupancy $d_{occ}$, (c) 
local and nonlocal spin correlations $\langle S_{1z} S_{iz} \rangle$, 
and (d) molecular orbital occupancies $n_m$ as functions of Coulomb energy 
for $T=0.005$. Orbitals 3 and 4 are doubly degenerate. There is no 
indication of hysteresis behavior in the critical region $U=3\ldots4$.
}\label{nvsu}\end{figure}

These results suggest that non-local correlations in the honeycomb lattice 
induce a paramagnetic semi-metal to insulator Mott transition in the range 
$U=3\ldots4$. Because of the continuous nature of the transition (see below), 
it is difficult to identify the precise value of the critical interaction.
Nevertheless, our finding is consistent with the variational QMC simulations 
\cite{meng} and the QMC/DMFT calculations\cite{wu} which yield $U_c\approx 3.6$.   
It is also in qualitative agreement with earlier finite-size cluster QMC
simulations which gave $U_c\approx 4.5$ (Ref.~\onlinecite{sorella}) and 
$U_c\approx 4\ldots5$  (Ref.~\onlinecite{paiva}). On the other hand, all 
of these values are significantly lower than the ones obtained within 
single-site DMFT which yields $U_c\approx 10\ldots 13$.\cite{jafari,tran} 
Moreover, in agreement with Refs.~\onlinecite{sorella,paiva,meng,wu}
we do not find any hysteresis behavior for increasing versus decreasing 
$U$, as shown in Figure~3(b) for the double occupancy, 
indicating that the transition is continuous. In contrast, within local
DMFT the transition was shown to be of first order.\cite{tran}      
Fig.~3(c) shows the onsite and intersite spin correlations, 
$\langle S_{1z} S_{iz} \rangle$, for $i=1\ldots4$. The onsite and second 
neighbor components are positive, while the first and third neighbor 
components are negative, underlining the antiferromagnetic nature of the 
spin correlations.

One of the interesting effects of Coulomb interactions in multi-orbital 
systems is the possibility of correlation-induced charge transfer between 
orbitals. As shown in Figure 1, the six-site unit cell of the honeycomb
lattice maybe viewed as consisting of six molecular orbitals which are split 
by an effective crystal field and therefore have different orbital occupancies.    
Figure 3(d) shows the variation of these occupancies with Coulomb energy.
Evidently, there is little orbital polarization, a result that was also 
observed in CDMFT calculations for the square and triangular lattices.
\cite{al2008} Moreover, the double occupancy, the spin correlations, and
the orbital occupancies reveal no clear sign of a Mott transition in the 
region where the spectral distribution exhibits the opening of a gap.

\begin{figure}  [t!] 
\begin{center}
\includegraphics[width=4.0cm,height=6.5cm,angle=-90]{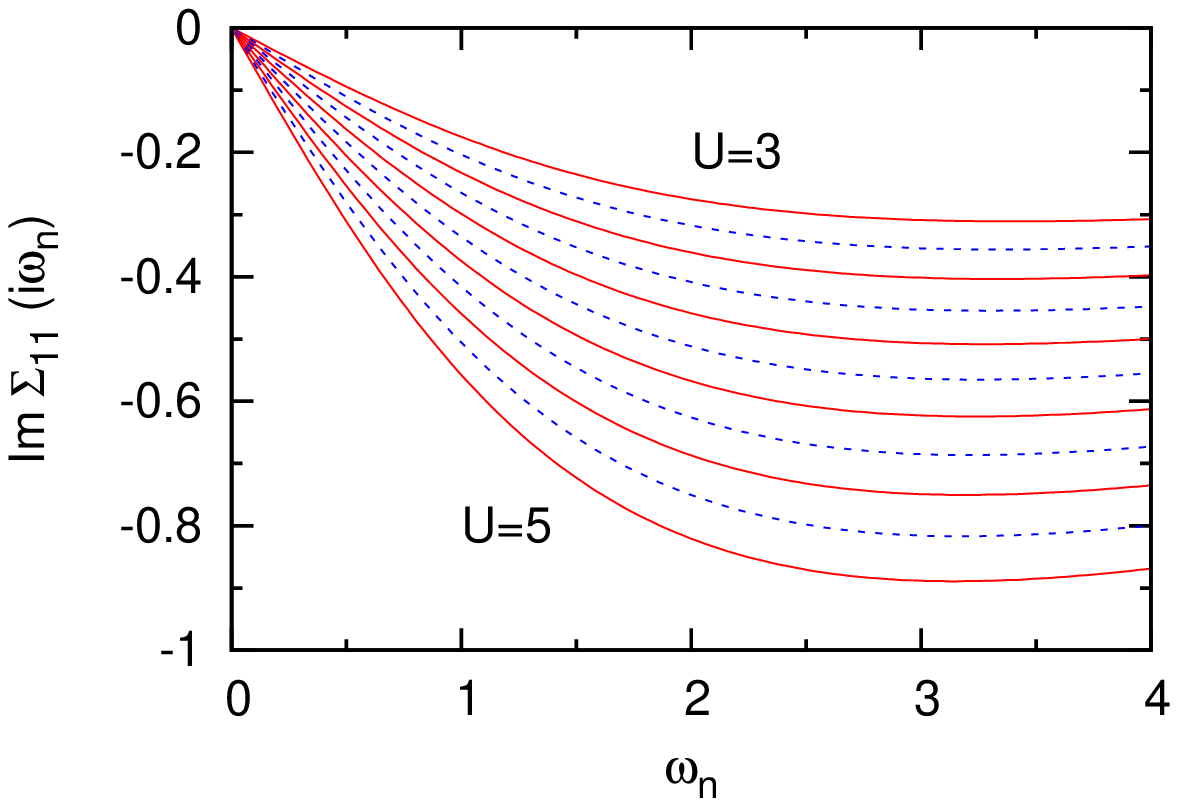}
\end{center}
\vskip-7mm \ \ \ (a)\hfill \mbox{\hskip5mm}\vskip-3mm
\begin{center}
\includegraphics[width=4.0cm,height=6.5cm,angle=-90]{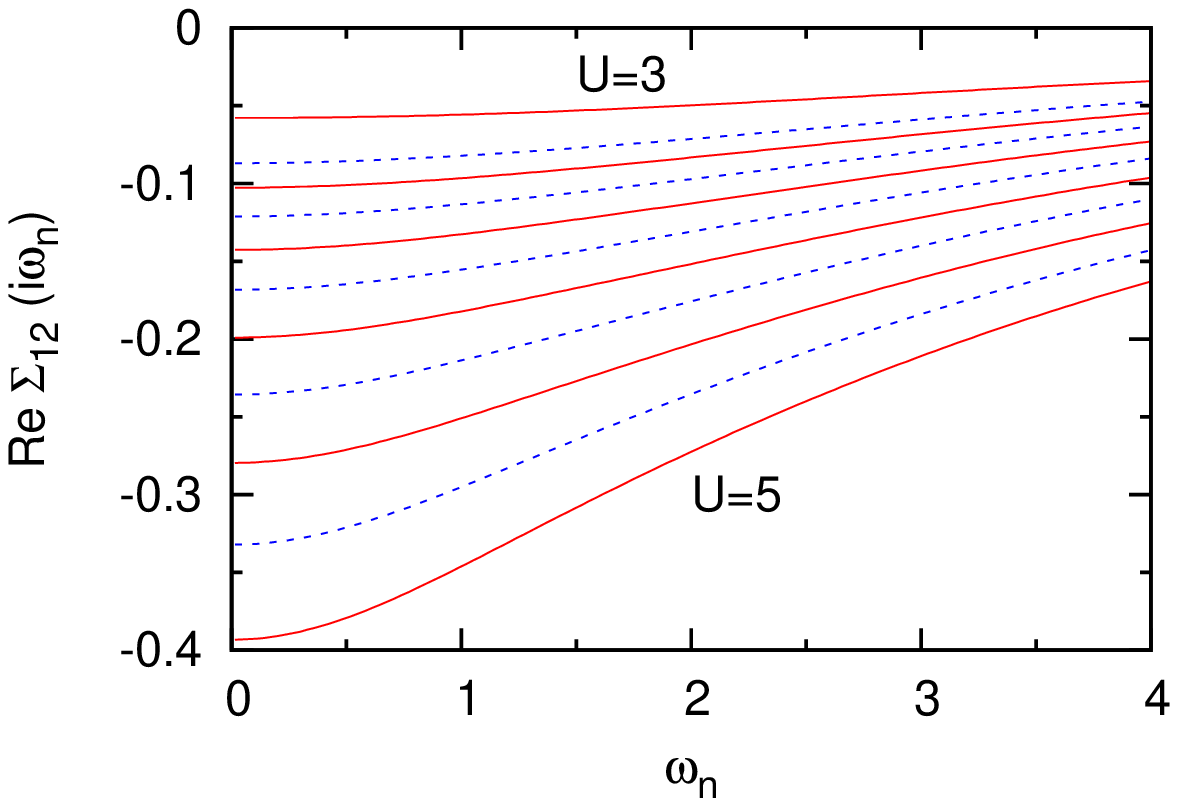}
\end{center}
\vskip-7mm \ \ \ (b)\hfill \mbox{\hskip5mm}\vskip-3mm
\begin{center}
\includegraphics[width=4.0cm,height=6.5cm,angle=-90]{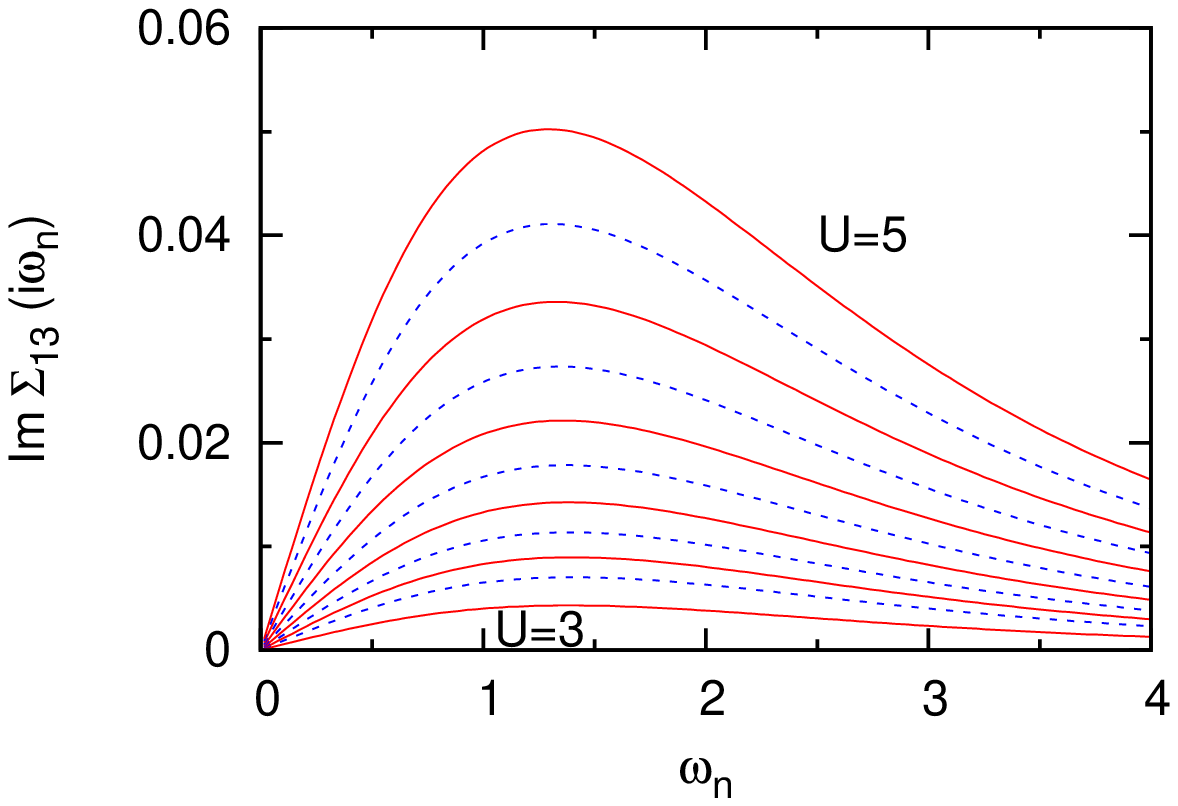}
\end{center}
\vskip-7mm \ \ \ (c)\hfill \mbox{\hskip5mm}\vskip-3mm
\begin{center}
\includegraphics[width=4.0cm,height=6.5cm,angle=-90]{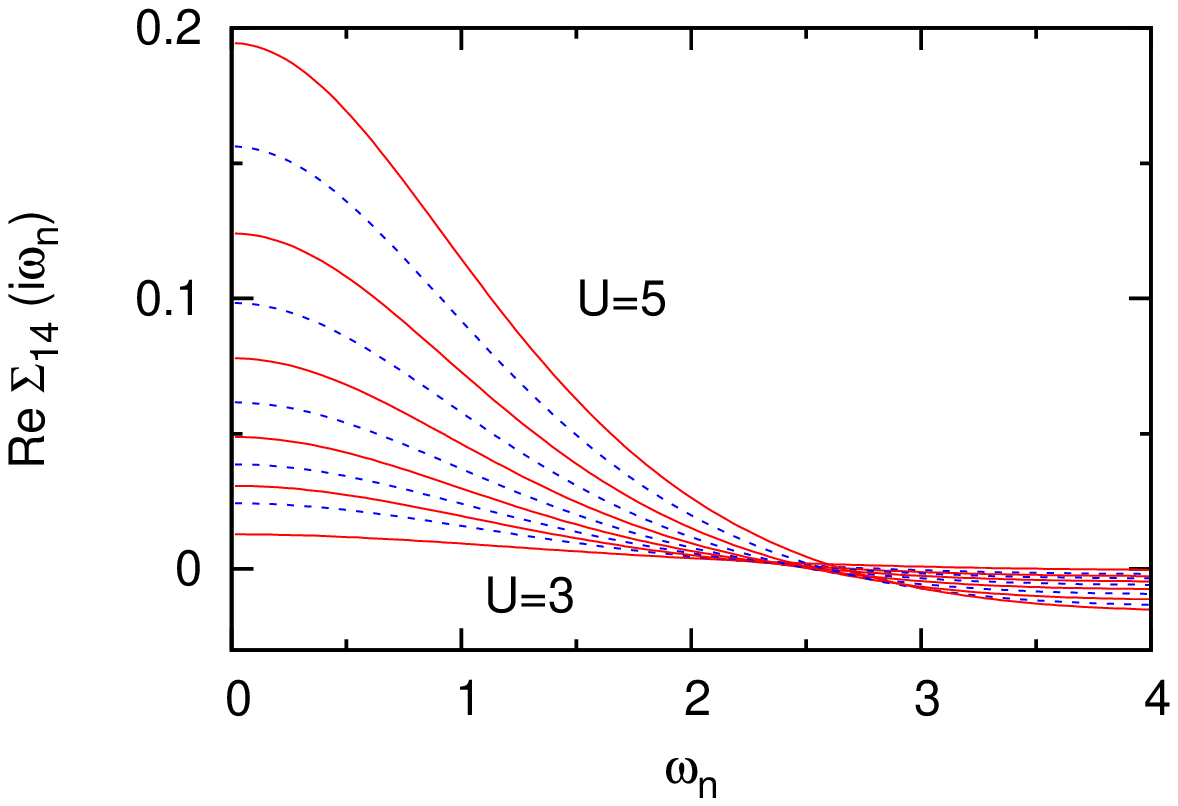}
\end{center}\vskip-3mm
\vskip-7mm \ \ \ (d)\hfill \mbox{\hskip5mm}
\caption{(Color online)
Self-energy components $\Sigma_{1i}(i\omega_n)$, $i=1\dots4$, for honeycomb 
lattice in cluster site representation as functions of Matsubara frequency 
for Coulomb energies $U=3\ldots5$ in steps of $0.2$; $T=0.005$. 
}\label{Sigma}\end{figure}

To analyze the nature of the semi-metal to insulator transition, it is 
therefore necessary to examine the non-local contributions to the self-energy.    
Figure~4 shows the four independent components of the cluster self-energy 
$\Sigma(i\omega_n)$ within the site basis, for Coulomb energies in the 
region of interest, $U=3\ldots5$. For symmetry reasons, 
$\Sigma_{11}(i\omega_n)$ and $\Sigma_{13}(i\omega_n)$ are purely imaginary. 
They behave as $\sim i\omega_n$ at low frequencies. In contrast,  
$\Sigma_{12}(i\omega_n)$ and $\Sigma_{14}(i\omega_n)$ are real and approach 
a finite value in the limit $\omega_n=0$.

In a seminal paper long before the synthesis of graphene, Gonz\'alez {\it et al.}
\cite{gonzalez} studied the influence of electron-electron interactions 
on the quasiparticle lifetime in a single layer of graphite. Taking into
account the long-range nature of the Coulomb interaction, their renormalization
group calculations indicate that the suppression of electronic screening at 
low frequencies yields deviations from conventional Fermi-liquid behavior, 
with Im\,$\Sigma(\omega)$ approximately linear in $\omega$ for 
$\omega=0.4\ldots3$~eV (for $t=2.4$~eV).
  
To determine possible non-Fermi-liquid contributions to the self-energy
derived within the present ED/CDMFT approach, we have carefully searched 
for $\omega_n {\rm ln}(\omega_n)$ behavior in the imaginary components 
$\Sigma_{11}(i\omega_n)$ and $\Sigma_{13}(i\omega_n)$. Within the accuracy 
of our results, these functions do not indicate any such deviations and seem 
to be well proportional to $i\omega_n$ in the entire range $U=0\ldots5$.
Also, they do not indicate a finite limiting value for 
$\omega_n\rightarrow 0$ which would imply a finite lifetime for states
near the Fermi energy. Thus, the non-Fermi-liquid properties obtained in
Ref.~\onlinecite{gonzalez} seem to be associated with the long-range part of the
Coulomb repulsion which is absent in the Hubbard model for purely onsite
interactions.\cite{guinea}  

\begin{figure}  [t!] 
\begin{center}
\includegraphics[width=4.5cm,height=6.5cm,angle=-90]{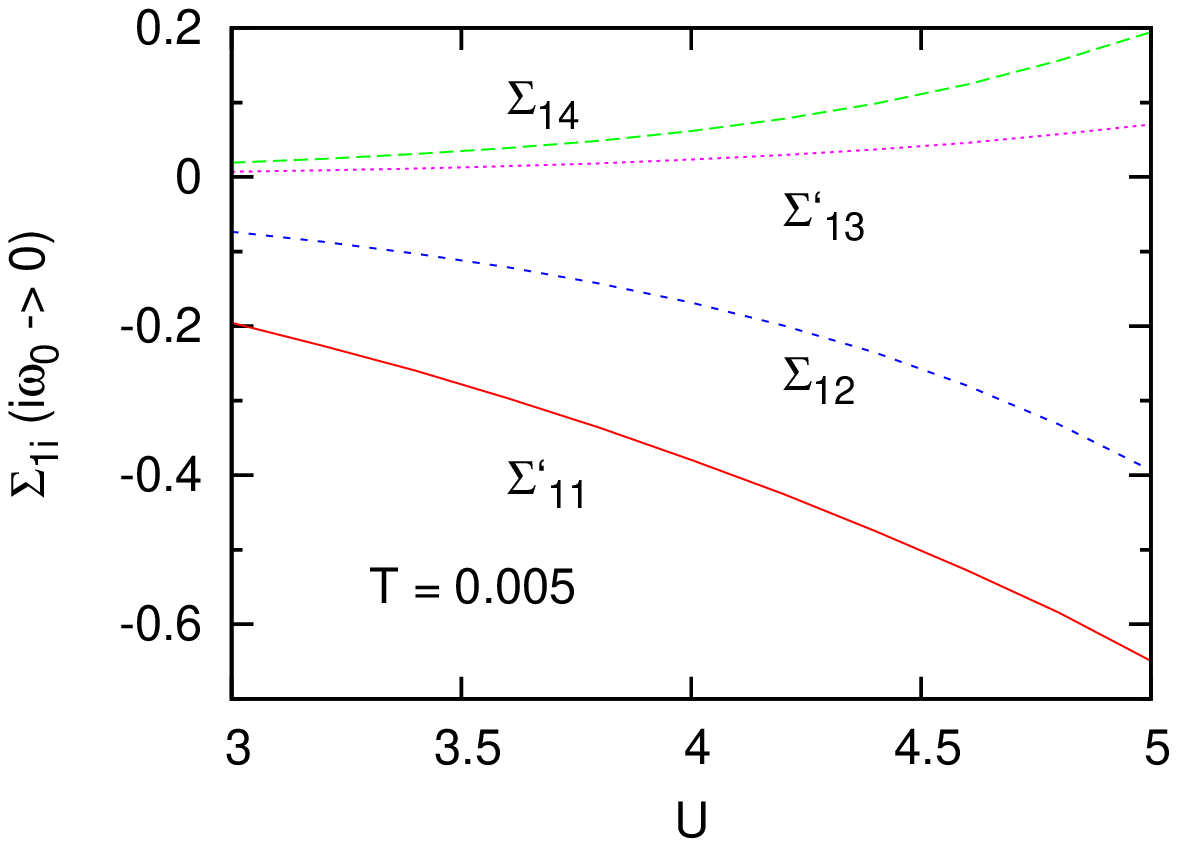}
\end{center}
\vskip-7mm \ \ \ (a)\hfill \mbox{\hskip5mm}\vskip-3mm
\begin{center}
\includegraphics[width=4.5cm,height=6.5cm,angle=-90]{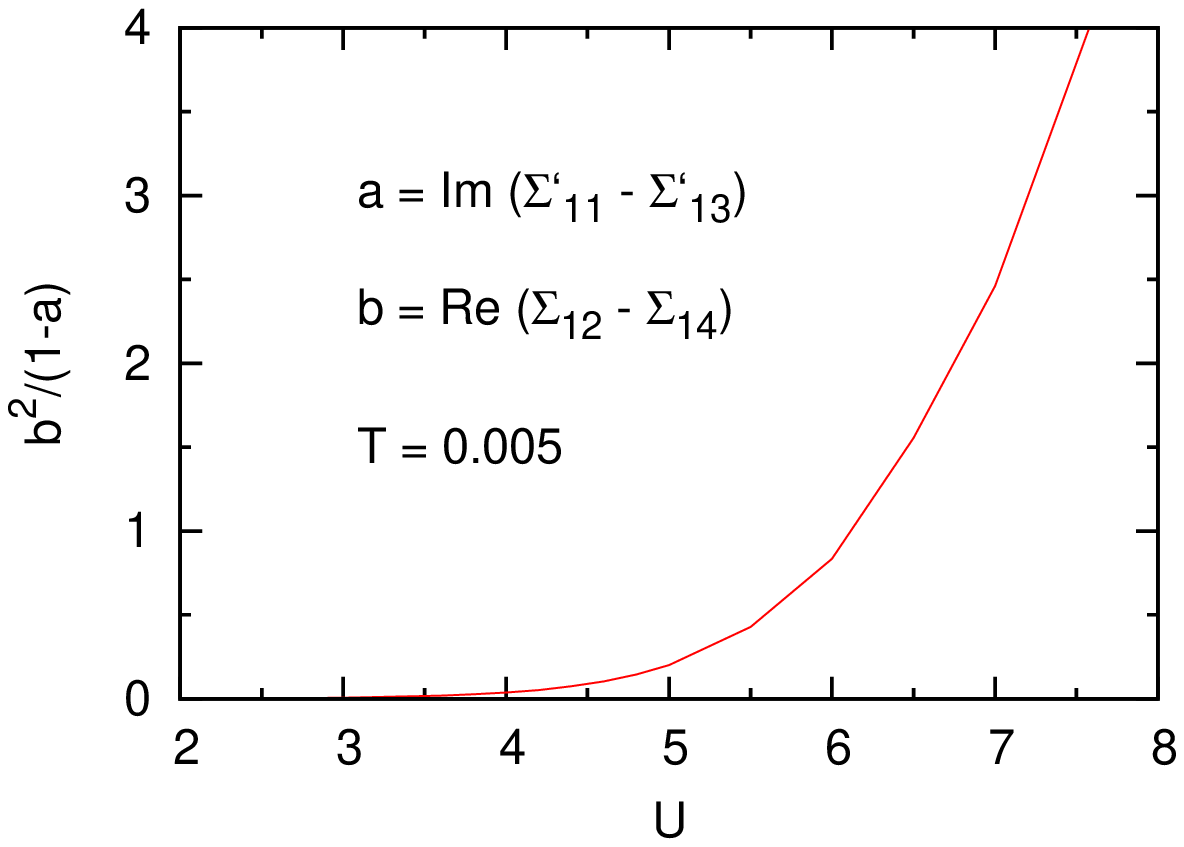}
\end{center}\vskip-3mm
\vskip-7mm \ \ \ (b)\hfill \mbox{\hskip5mm}
\caption{(Color online)
(a) Low-frequency limits of self-energy components: slopes of 
Im\,$\Sigma_{11}(i\omega_n)$, Im\,$\Sigma_{13}(i\omega_n)$, and values of 
Re\,$\Sigma_{12}(i\omega_n)$, Re\,$\Sigma_{14}(i\omega_n)$, as functions of $U$.
(b) Amplitude $b^2/(1-a)$ of insulating contribution to $\Sigma(K,i\omega_n)$,
Eq.~(\ref{SK}), as a function of $U$.  $T=0.005$
}\label{Sigmavsu}\end{figure}

We note, however, that to understand the spectral features of the quasiparticle
density of states, it is not sufficient to study the self-energy components 
shown in Fig.~4. In particular, these isolated components do not provide 
any evidence for a Mott transition in the region $U=3\ldots4$, where the 
density of states shown in Fig.~2 indicates the opening of a gap. 
To illustrate the smoothness of the self-energy 
components in this range of Coulomb energies, we show in Fig.~5(a) 
the slopes of Im\,$\Sigma_{11}$ and Im\,$\Sigma_{13}$, and the values
of Re\,$\Sigma_{12}$ and Re\,$\Sigma_{14}$ in the low-frequency limit.
Evidently, these individual components do not reveal the
existence of the Mott transition seen in the density of states.
This behavior differs qualitatively from the Hubbard model for the square
lattice at half-filling, where at the metal insulator transition the $(\pi,0)$ 
component of the self-energy changes from $\sim i\omega_n$ to $\sim 1/i\omega_n$ 
at small $\omega_n$, and the real part of the $(0,0)$ and $(\pi,\pi)$ 
components exhibits a jump.\cite{zhang,park}  

The origin of this apparent discrepancy is the fact that, as pointed out 
above, the local interacting density of states depends in a highly nonlinear 
manner on all non-local self-energy components $\Sigma_{ij}$. 
This is evident from the expression 
for the lattice Green's function, Eq.~(\ref{G}), where the hopping matrix 
$t({\bf k})$ and $\Sigma(i\omega_n)$ cannot be simultaneously diagonalized, 
as indicated also in Eq.~(\ref{Gm}). To account for this admixture of intersite 
self-energy elements, it is useful to examine the $6\times6$ cumulant matrix
\begin{equation}
       M(i\omega_n) = [i\omega_n - \Sigma(i\omega_n)]^{-1}. 
\label{M}
\end{equation}
Since $M$ has the same symmetry properties as $\Sigma$, its nonlocal 
components are given by 
\begin{eqnarray}
  M_{11} &=& [(M_1+M_2) + 2(M_3+M_4)]/6 \nonumber \\
  M_{12} &=& [(M_1-M_2) -  (M_3-M_4)]/6 \nonumber \\
  M_{13} &=& [(M_1+M_2) -  (M_3+M_4)]/6 \nonumber \\
  M_{14} &=& [(M_1-M_2) + 2(M_3-M_4)]/6 
\label{Mij}
\end{eqnarray}
where the diagonal molecular orbital elements are  
\begin{equation}
       M_m(i\omega_n) = [i\omega_n - \Sigma_m(i\omega_n)]^{-1}. 
\label{Mm}
\end{equation}

\begin{figure}  [t!] 
\begin{center}
\includegraphics[width=4.5cm,height=6.5cm,angle=-90]{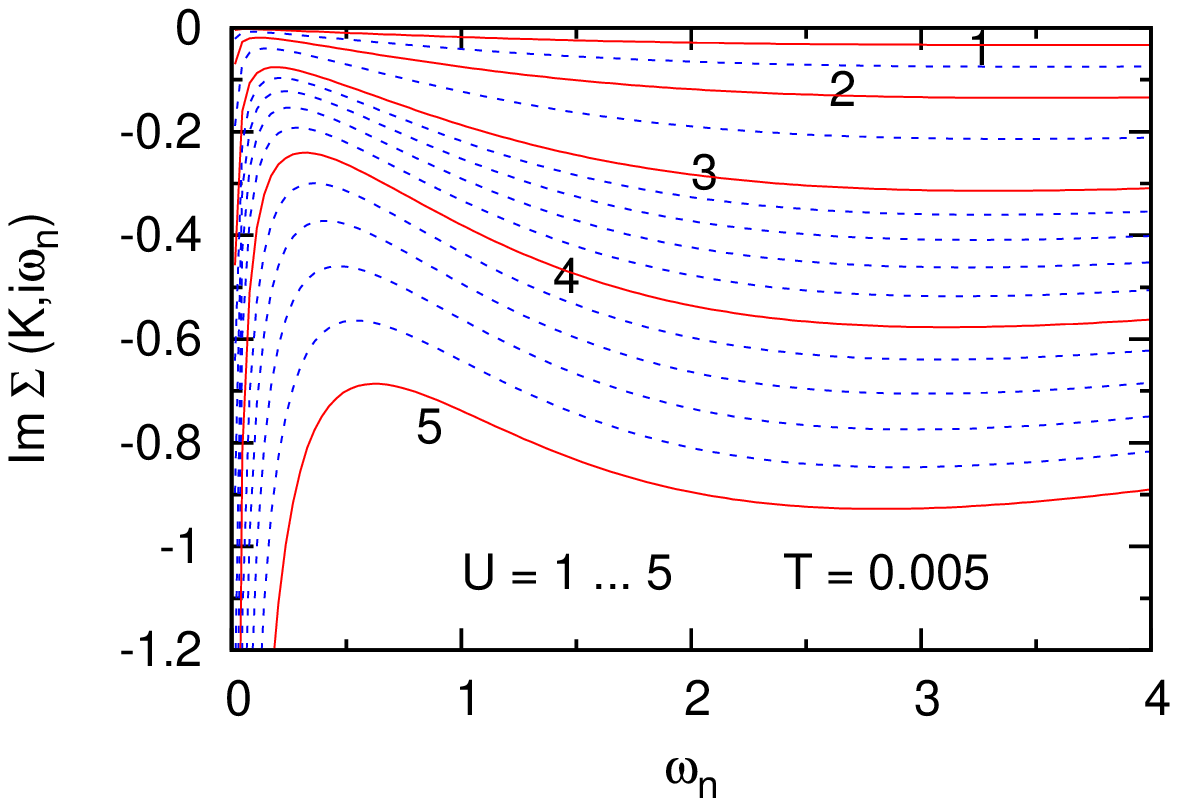}
\end{center}
\vskip-7mm \ \ \ (a)\hfill \mbox{\hskip5mm}\vskip-3mm
\begin{center}
\includegraphics[width=4.5cm,height=6.5cm,angle=-90]{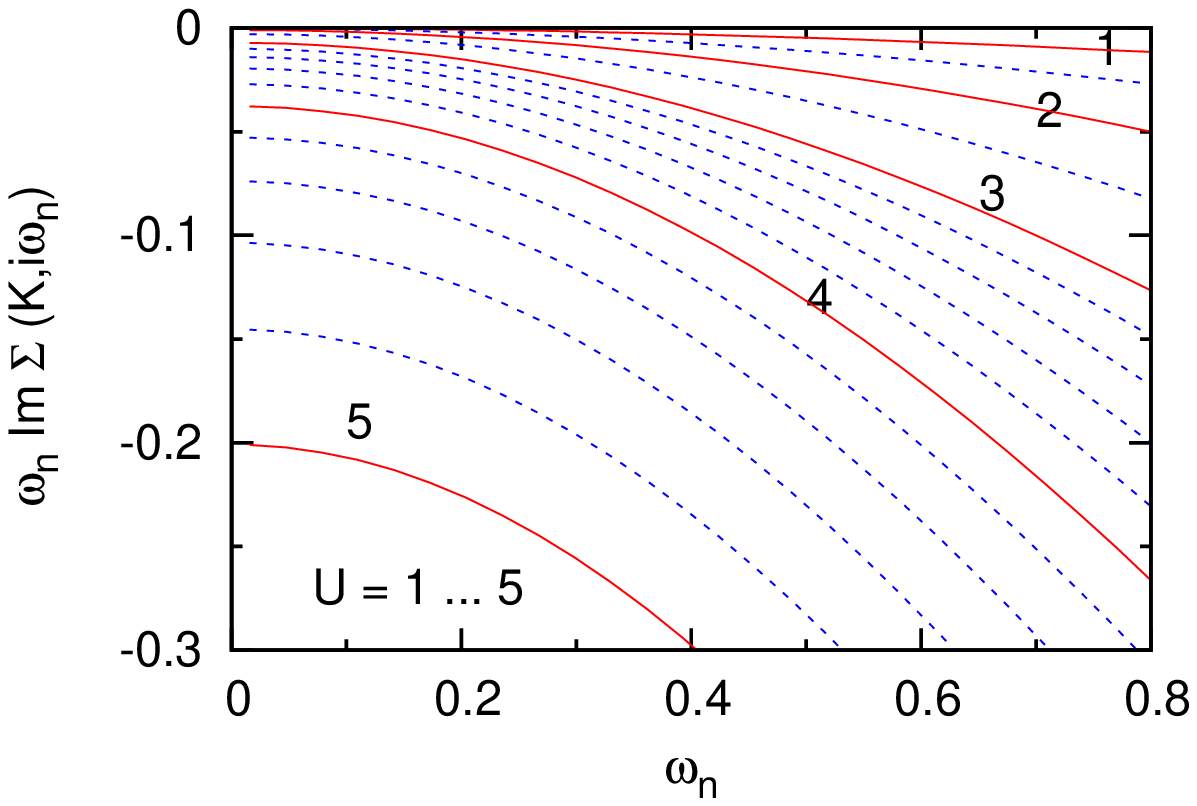}
\end{center}
\vskip-7mm \ \ \ (b)\hfill \mbox{\hskip5mm}
\caption{(Color online)
(a) Imaginary part of self-energy at $K$, Eq.~(\ref{SK}), as a function 
of Matsubara frequency, for $U=1\ldots5$ at $T=0.005$.
(b) $\omega_n {\rm Im}\,\Sigma(K,i\omega_n)$, demonstrating the range of 
the insulating part of the self-energy at Dirac points $K$.
Solid red curves denote integer values of $U$, dashed blue curves denote 
intermediate values of.    
}\label{SigmaK}\end{figure}

The opening of the Mott gap takes place at the six $K$ points of the Brillouin 
Zone. To analyze the behavior of the cumulant at these points, we make use of 
the periodization \cite{parcollet}
\begin{equation}
 M({\bf k},i\omega_n)=\frac{1}{6}\sum_{ij=1}^6 e^{i{\bf k}\cdot({\bf a}_i-{\bf a}_j)} 
           M_{ij}(i\omega_n),    
\label{Mlat}
\end{equation}
where ${\bf a}_i$ are the positions within the six-site cluster. 
At $K=2\pi(2/3\sqrt3,0)$ and $K'=2\pi(1/3\sqrt3,1/3)$, this expression 
simplifies to  
\begin{eqnarray}
 M(K,i\omega_n) &=& M_{11}(i\omega_n) - M_{13}(i\omega_n)  \nonumber\\
                &=& [M_3(i\omega_n) + M_4(i\omega_n)]/2 .  
\label{MK}
\end{eqnarray}
The self-energy at $K$ is therefore given by  
\begin{eqnarray}
  \Sigma(K,i\omega_n) &=& i\omega_n - M^{-1}(K,i\omega_n) \nonumber\\
                      &\approx&  i\omega_na + \frac{b^2}{i\omega_n(1-a)}, 
                   \hskip3mm  \omega_n\rightarrow0, 
\label{SK}
\end{eqnarray}
where $a$ is the initial slope of 
Im\,$[\Sigma_{11}(i\omega_n)-\Sigma_{13}(i\omega_n)]$ 
and $b$ the low-frequency limit of 
Re\,$[\Sigma_{12}(i\omega_n)-\Sigma_{14}(i\omega_n)]$.
This self-energy  is shown in Figure 6 for various Coulomb energies.
The above expression indicates that $\Sigma(K,i\omega_n)$ is imaginary
as expected for particle-hole symmetry at the Dirac points. It consists of 
metallic ($\sim i\omega_n$) and insulating ($\sim 1/i\omega_n$) contributions.
The insulating term, which is responsible for the opening of the Mott gap,
increases quadratically with $b={\rm Re}\,[\Sigma_{12}-\Sigma_{14}]$.
Thus, the semi-metal to insulator transition is driven primarily by
the nearest-neighbor component of the non-local self-energy, with a minor
additional contribution due to the third-neighbor self-energy, and a weak
renormalization related to the initial slope of 
Im\,$[\Sigma_{11}(i\omega_n)-\Sigma_{13}(i\omega_n)]$.
The variation of the amplitude $b^2/(1-a)$ of the insulating term with
Coulomb energy is depicted in Fig.~5(b). The comparison with Fig.~3(a)
demonstrates that the excitation gap $\Delta$ roughly tracks the amplitude 
of this term. 

According to the results shown in Figure~4, $a\approx -0.2\ldots {-0.6}$ 
and $b\approx -0.05\ldots -0.6$ in the range $U=3\ldots5$.
Thus at $U=3$ the amplitude of the insulating term is about $10^2$ times 
smaller than at $U=5$. Nevertheless, this small contribution is responsible
for the pseudogap below the Mott transition, indicating the breakdown of 
Fermi-liquid behavior in the metallic phase.
For $U<2$ we find $\vert b\vert < \omega_0$, so that the pseudogap can 
no longer be resolved within the accuracy of ED. A similar pseudogap induced 
by short-range correlations at half-filling was observed below the Mott 
transition in the Hubbard model for the square lattice.\cite{kyung,zhang,park} 
Neglecting the small insulating term of $\Sigma(K,i\omega_n)$ sufficiently 
far below the transition, the effective mass enhancement of the quasiparticle
bands near the Dirac points is given by $m^*/m = 1-a = 1.0\ldots1.2$ for 
$U=0\ldots 3$, i.e., it does not diverge at the Mott transition, in contrast 
to results derived within single-site DMFT. This finding is also consistent
with the behavior seen on the square lattice.\cite{zhang}

\section{Summary}

The influence of onsite Coulomb interactions on the electronic properties
of the honeycomb lattice has been investigated within cluster dynamical 
mean field theory combined with exact diagonalization. The interacting 
density of states exhibits the opening of a Mott gap in the region 
$U=3\ldots 4$, which is caused by a change of the self-energy at the 
Dirac points of the Brillouin Zone from metallic to insulating behavior. 
This transition is in good agreement with finite-size extrapolations of 
variational QMC simulations and with continuous-time QMC calculations 
based on cluster DMFT. As a result of short-range fluctuations, the 
critical Coulomb energy is significantly smaller than in single-site 
DMFT calculations. Also, a narrow pseudogap is found close to the Mott 
transition. Finally, the effective mass shows a moderate enhancement at
finite $U$, but it does not diverge at the transition. 
 
The consistency between the ED and QMC calculations for the honeycomb 
lattice, including the variation of the Mott gap with onsite Coulomb 
repulsion, suggests that, as long as the overall size of the Hilbert space 
is sufficiently large, yielding small enough level spacing, the use of 
only one bath level per impurity orbital can be adequate. This situation
differs from the one for fewer sites or orbitals, where more bath levels 
per impurity level must be included to achieve sufficiently large Hilbert 
spaces. 

\bigskip

{\bf Acknowledgements}\ \ I like to thank Hiroshi Ishida for valuable discussions.
The ED/CDMFT calculations were carried out on the J\"ulich Juropa computer.

\end{document}